\documentclass{aa}  

\usepackage{graphicx}
\usepackage{txfonts}
\usepackage{hyperref}
\usepackage{amsmath,amssymb}  
\begin{document}

   \title{MUSE observations of V1425 Aql reveal an arc-shaped nova shell.}

   \subtitle{}

   \author{Lientur Celed\'on\inst{1}
   \and
   Claus Tappert\inst{1}
   \and
   Linda Schmidtobreick\inst{2}
   \and
   Fernando J. Selman\inst{2}          
   }

   \institute{
            {Instituto de F\'isica y Astronom\'ia, Universidad de Valpara\'iso, Valpara\'iso, Chile
            }
            \and
            {European Southern Observatory, Casilla 19001,
              Santiago 19, Chile}\\  
            \email{lientur.celedon@postgrado.uv.cl}
            }

   \date{Received xxx; accepted yyy}

 
  \abstract
   {Nova shells are the remnants of a nova eruption in a cataclysmic variable system. By studying their geometry we can better understand the physical mechanisms that shape them during the nova eruption. A nova shell that challenges our current understanding of these processes is the shell observed around V1425 Aql. It has at least two different components: an inner and symmetric shell, and an outer and asymmetric shell, with the latter expanding faster than the former. The physical reason behind the asymmetric ejecta is not clear.}
   {We aim to characterize the properties and differences between these two components to understand the origin behind the unusual shape.}
   {We acquired MUSE data to study the spatial position and kinematics of the expanding gas across the shell. Our analysis includes channel maps, position-velocity diagrams, and the reconstruction of the 3D geometry of the nova shell.}
   {Several emission lines are detected within the MUSE wavelength coverage, including but not limited to Balmer, Oxygen, Nitrogen, and Helium lines.
   There are significant differences in the spectra of the inner and outer shells, with the latter being observed only in forbidden transitions, while the inner shows a mix of forbidden and allowed ones.
   Our analysis reveals that the outer shell has a geometry consistent with an arc-shaped structure that partially encircles the more spherical inner shell. Within the inner shell, clumpy structures start to be noticeable in the lines of H$\rm\alpha$+[N{\sc ii}].
   }
   {We have constrained the geometry of the outer shell to an arc-shaped structure, although the physical reason behind its origin is still eluding us. Further monitoring of the evolution of both shells in this object might help to clarify the mechanism behind this unusual configuration.}


   \maketitle
%

\section{Introduction}

Cataclysmic variables (CVs) are a group of binary systems where a low-mass main sequence star transfers mass to its white dwarf (WD) companion via Roche lobe overflow \citep[See][for a review on CVs]{Warner1995cvs..book.....W}.
The accreted material is eventually deposited over the WD surface via an accretion disk, or directly funneled onto the WD poles in the case of magnetic WDs.
Whatever the case may be, once a critical amount has been accreted onto the WD's surface the pressure at its bottom layers causes the Hydrogen to suddenly fuse into Helium, producing a thermonuclear runaway process \citep{Starrfield+2016PASP..128e1001S}. This is the beginning of a nova eruption, a transient phenomenon in CVs that can last days to months. They are a recurrent process in the lifetime of CVs with recurrence time scales of thousands of years \citep{Schmidtobreick+2015MNRAS.449.2215S}, playing a fundamental role in their evolution.
One of its consequences is the ejection to the interstellar medium (ISM) of the material that has been accreted into the WD surface. The mass of the ejecta, $M_{\mathrm{ej}}$, is in the order of 10$^{-5}$-10$^{-4}$ M$_\odot$, and it can reach velocities of hundreds, even thousands of km s$^{-1}$ \citep[e.g.][]{Gehrz+1998PASP..110....3G, Aydi+2020ApJ...905...62A}. The expanding material forms a nebular structure around the system called a nova shell. They are an important piece of the puzzle which is the understanding of the physical processes behind a nova eruption.

Nova shells tend to deviate from a purely spherical shape and instead show more complex geometries like prolate shapes, equatorial rings, polar filaments or clumpiness \citep[e.g.][]{Gill&O'Brien1998MNRAS.300..221G, Downes&Duerbeck2000AJ....120.2007D, Santamaria+2022MNRAS.512.2003S, Celedon+2024A&A...681A.106C, Santamaria+2024MNRAS.530.4531S}.
Although these structures are only revealed to us once the nova shell has expanded enough to be resolved, the process that shaped it must be found at earlier stages, during the nova eruption itself.
It has been proposed that during a nova eruption, the main ejection event is followed by several minor ejections of material in time scales of days. This idea of multiple ejections is able to explain the spectral features observed in the days after the eruption, as well as the multi-wavelength observations in X-ray and radio, which are supposed to have originated from the shocks generated when these multiple ejections collide with each other \citep{Aydi+2020ApJ...905...62A, Chomiuk+2021ApJS..257...49C}.
As a prediction of this model, all nova shells should have ellipsoidal geometries, and while many of them present this morphology, those that do not can be studied to help us to better understand the physics behind a nova eruption.
An obvious explanation for the deviation of the expected morphology is the interaction between the expanding shell and the surrounding ISM. This is expected to be an important factor in old ($\sim$ centuries) nova shells, as it has been observed that they can expand freely through the ISM during at least a century \citep{Santamaria+2020ApJ...892...60S, Celedon+2024A&A...681A.106C}.

One prime example of a nova shell that deviates from the usual is the one around V1425 Aql.
The nova eruption of V1425 Aql was first observed on February 2, 1995, reaching a maximum brightness of 8 mag \citep{Nakano+1995IAUC.6133....1N}.
A light curve analysis of the system shortly after the nova eruption shows two strong periodicities, one at 6.14 h, and a second at 1.44 h \citep{Retter+1998MNRAS.293..145R}. The authors associated the first periodicity to the orbital motion, and the second to the spin of the WD, implying that the system is likely an intermediate polar: a CV hosting a weak magnetic WD.
However, a subsequent study by \citet{Worpel+2020A&A...639A..17W} was not able to detect any X-ray emission coming from the system, questioning the possibly magnetic nature of the WD.
Spectroscopic observations following the nova eruption reveal the presence of thin dust material, a dust-to-gas ratio $<$ 10$^{-3}$, evidence for clumpy structures within the shell, highly enhanced C, N, and O with respect to solar abundance, and an ejected mass in the order of 2-4$\times$10$^{-5}$ M$_\odot$ \citep{Mason+1996ApJ...470..577M, Kamath+1997AJ....114.2671K, Lyke+2001AJ....122.3305L}.

\begin{figure}
    \centering
    \includegraphics[width=1.0\columnwidth]{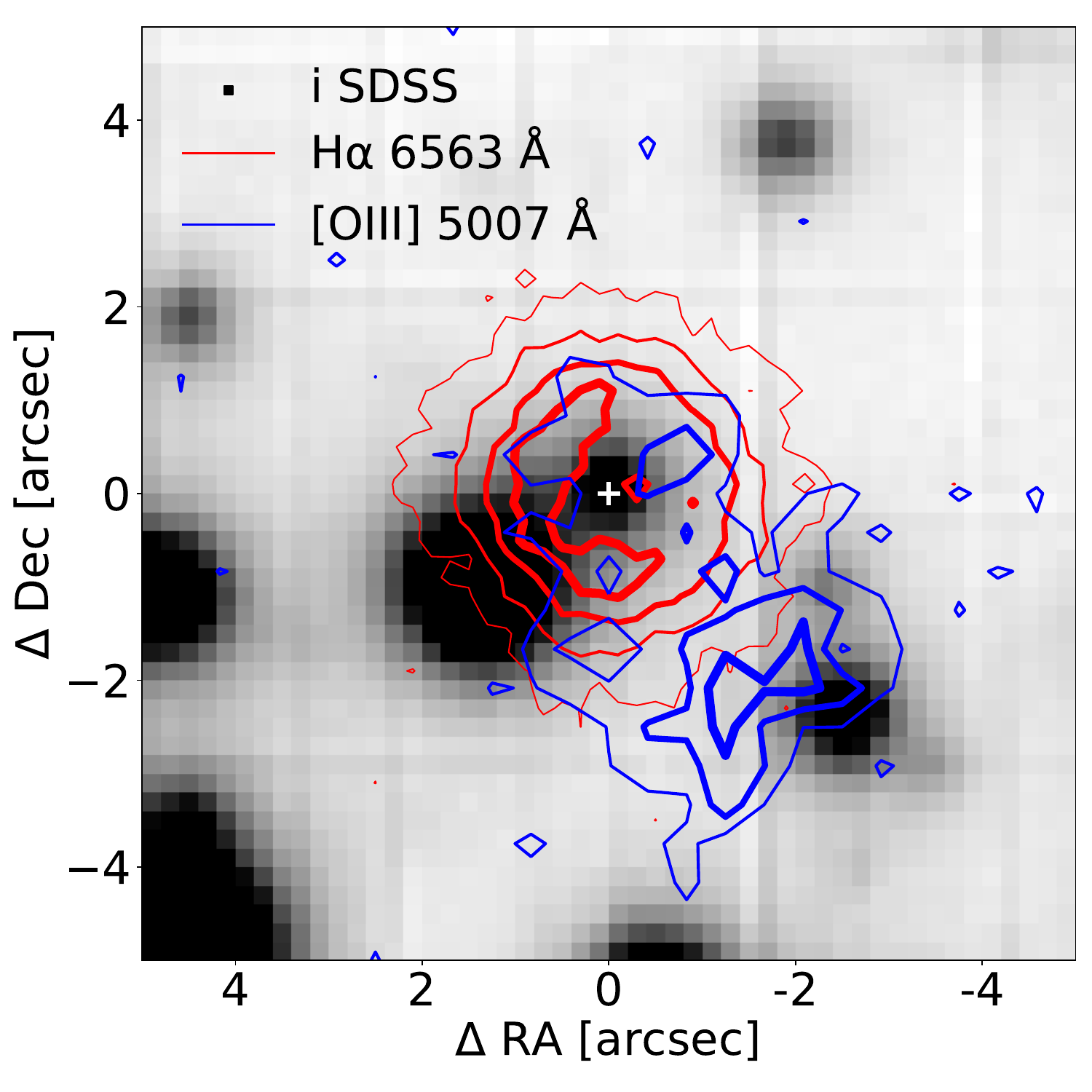}
    \caption{Bandpass image of 10$\times$10 arcsecs$^2$ centred at the position of V1425 Aql, showing the extension of the nova shell. The grey image corresponds to an i\_SDSS image, while the contours show the extension of the inner and outer shell as seen at the H$\rm\alpha$ (red) and [O{\sc iii}]\,$\rm\lambda$\,5007\,{\AA} (blue) wavelengths, respectively, with thicker lines denoting regions with higher emissivity.
    Both the image and the contours were obtained from the MUSE datacube.
    The difference in spatial extension between both shells is evident, as well as the asymmetry observed in the outer shell.
    }
    \label{fig:shell1}
\end{figure}

This CV and its expanding shell received little attention until recently, when \citet{Tappert+2023A&A...679A..40T}, T23 from here on, presented new Southern Gemini Multi-Object Spectrograph (GMOS-S) long-slit spectroscopic (LSS) and narrow-band (NB) photometric data of the system revealing the expanding material to consist of apparently two distinct shells: an inner and outer shell.
The inner shell is traced mainly by allowed transitions (Balmer, He{\sc i} and He{\sc ii}, and N{\sc ii}) but also in the forbidden lines of [O{\sc iii}] and [N{\sc ii}], while the outer shell is observed only in the forbidden lines of [O{\sc iii}] and [N{\sc ii}].
What calls the most attention about this nova shell is the evident differences in the spatial position between both shells, with the inner one being symmetrically centred on the position of the binary, and the outer one consisting of a cone of material expanding to a larger distance from the binary and exclusively in the south-west direction (Fig.~\ref{fig:shell1}).
The larger extension of the asymmetric shell suggests a higher expansion velocity, which was confirmed by the LSS data: the velocities of the inner shell are $\sim$500 km s$^{-1}$, while the outer shell reaches up to $\sim$1500 km s$^{-1}$.
The differences observed in the [O{\sc iii}]/[N{\sc ii}] ratio between both shells, plus the absence of H in the outer shell suggest significant differences in the abundances, electron density and/or electron temperature between them. 

Both shells appear to have originated during the same event (the 1995 eruption), although the uncertainties allow the possibility for a time difference of up to a couple of years between the potentially individual ejections.
This kind of asymmetry has not been previously reported in any other shell of similar age and its origin is far from clear, with the authors not being able to discern between an intrinsic or extrinsic origin for it.
The peculiarities observed in this shell motivated us to continue its study using the Multi-Unit Spectrograph Explorer instrument. The characteristics of this instrument are ideal for the study of extended objects like nova shells \citep[e.g.][]{Celedon+2024A&A...681A.106C}, providing us with the capabilities to better characterize this nova shell and potentially clarify its origin.

\section{Data}

The Multi-Unit Spectrograph Explorer, MUSE, instrument is an integral field spectrograph installed at the Very Large Telescope observatory, Paranal, Chile, operated by the European Southern Observatory (ESO).
When observing in its standard wide-field mode configuration it provides a field of view of 1$\times$1 arcmin$^2$ with a sampling of 0.2$\times$0.2 arcsec$^2$ per pixel.
The wavelength range covers the optical part of the spectrum from 4800 to 9300 {\AA} with a wavelength resolution of 1.25 {\AA} per pixel \citep{Bacon+2010SPIE.7735E..08B}.
Its spectral resolution ($R\sim$2600 at H$\rm\alpha$) implies a resolution of $\sim$115 km\,s$^{-1}$ in the velocity domain, which is more than enough to resolve expanding structures with velocities of several hundreds of km\,s$^{-1}$, as it is the case of nova shells.

The acquisition of the MUSE data of V1425 Aql\footnote{ESO program ID: 105.20D8.001} consisted of four individual observations: three of them were collected on July 31, 2021, with a combined exposure time of 2500 seconds, and a fourth observation on August 5, 2021, consisting on 1200 s of exposure time.
The time differences between the MUSE data and the data published in T23 are $\sim$3 years for LSS data and $\sim$2 for the NB images. This corresponds to $\sim$10\% of the age of the shell at the time of MUSE observations.

The observations were carried out using the standard observation mode with adaptative optics to correct for the ground-based turbulence, thus improving the natural seeing of the image.
This, however, has the detriment that the wavelength region between 5800 and 5960 {\AA} is lost due to the saturation produced by the sodium lasers.
The four datacubes were reduced and then combined into a single one following the MUSE standard recipe \citep{Weilbacher+2020A&A...641A..28W}.
As a last step, the flux values stored in the datacube were corrected for extinction using a value of $E_\mathrm{(B-V)}=0.76$ mag \citep{Kamath+1997AJ....114.2671K, Tappert+2023A&A...679A..40T}, and the reddening law of Fitzpatrick \citep{Fitzpatrick1999PASP..111...63F} implemented in the $\tt PyNeb$ $\tt Python$ package \citep{Luridiana+2015A&A...573A..42L}. 

\section{Data analysis}

\subsection{A general view of the shell}

\begin{figure}
    \centering
    \includegraphics[width=1.0\columnwidth]{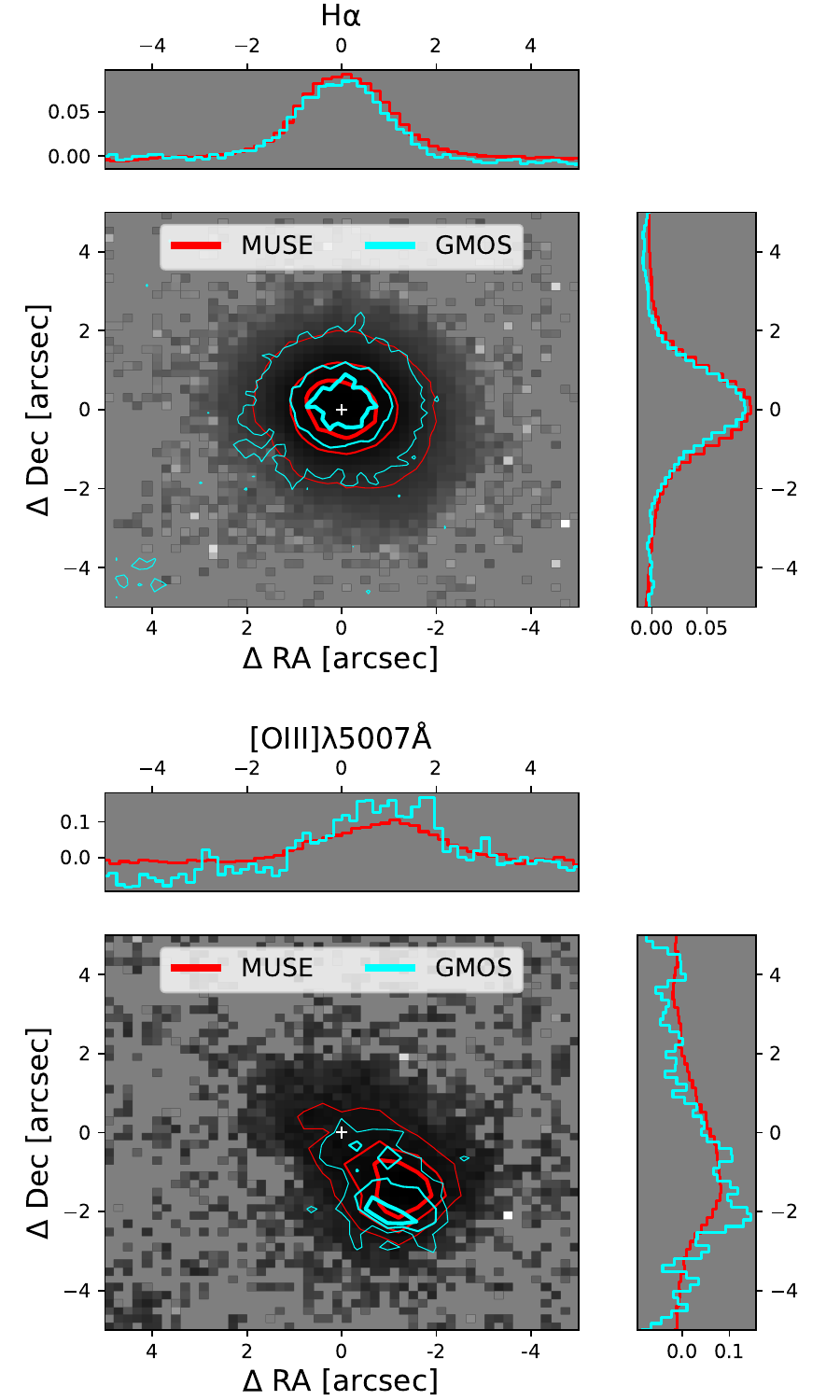}
    \caption{Comparison between the GMOS NB images presented in T23 and MUSE data for H$\rm\alpha$ (top) and [O{\sc iii}]\,$\rm\lambda$\,5007\,{\AA} (bottom). The MUSE images (grey background) were obtained by mimicking the bandpass used in the GMOS images. The contour levels indicate the same normalized fluxes for MUSE (red) and GMOS (cyan), while the white cross marks the position of the binary. The contours presented in H$\rm\alpha$ indicate no major differences between GMOS and MUSE, while [O{\sc iii}] show differences in the location of the brightest region within the outer shell, although they can be attributed to the noise in the GMOS data.}
    \label{fig:gmos_vs_muse}
\end{figure}

\begin{figure*}[h]
\centering
\includegraphics[width=1.0\textwidth]{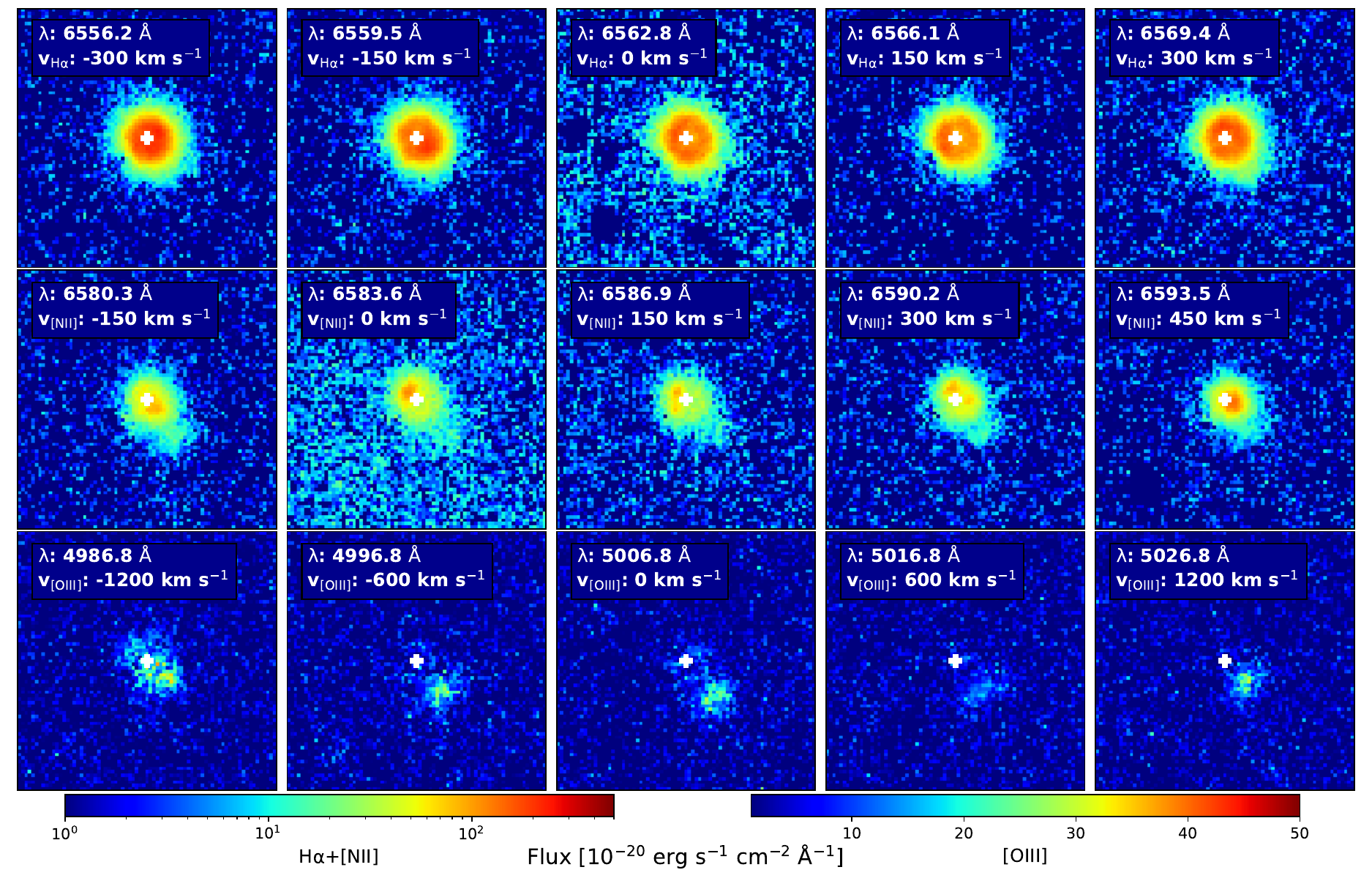}
\caption{Channel maps of V1425 Aql for the lines of H$\rm\alpha$ (top row), [N{\sc ii}]\,$\rm\lambda$\,6584\,{\AA} (middle row), and [O{\sc iii}]\,$\rm\lambda$\,5007\,{\AA} (bottom row).
Each channel map shows a region of 15$\times$15 arcsec$^{2}$. In all cases, the fluxes are presented in linear scales, with their ranges defined in the colour bars and the position of the binary marked with a white cross.
For each line, the velocities were chosen to illustrate the behaviour of the different lines.
The behaviour of the strongest lines in the nova shell as a function of their velocity illustrates the structures hidden within the inner shell, previously unnoticed in T23. 
}
\label{fig:channel_maps}
\end{figure*}

An overall view of the system is presented in Fig.~\ref{fig:shell1}, where a bandpass image corresponding to the i\_SDSS filter was created from the MUSE datacube using the available transmission for the GMOS-S telescope in the Spanish Virtual Observatory\footnote{\url{http://svo2.cab.inta-csic.es/theory/fps}}.
The contribution of the expanding shell in this filter is minimal if any, and therefore this emission defines the position of the binary.
This is located at the centre of the field and is found to be within a relatively crowded field. 
Two stars in the field are very close to the position of the binary, at $\sim$2 and $\sim$3 arcsecs, so that their projected positions overlap with the expanding ejecta (red and blue contours).
From the bandpass image, we determine the i\_SDSS magnitude of V1425 Aql to be 20.5$\pm$0.1 mag.

To show the extension of the inner shell we plot the contour levels at the rest wavelength of H$\rm\alpha$ after subtracting the flux of the continuum between 6620 and 6640\,{\AA}.
The contour levels are defined as 0.1, 0.25, 0.5 and 0.75 times the maximum of the observed flux in the continuum-subtracted image.
The projected geometry of the inner shell in the plane of the sky can be observed to be fairly circular, with a radius of $\sim$2 arcsecs.
The flux increases towards the centre of the shell, where clumpy material starts to appear as a region of higher flux in the east part of the shell.

The extension of the outer shell is shown by the emission of [O{\sc iii}] at 5007\,{\AA} with blue contours.
Similar to the H$\rm\alpha$ contours, these were obtained after subtracting the continuum corresponding to the wavelength range between 5040 and 5050\,{\AA}, and the levels indicate 0.25, 0.5 and 0.75 times the maximum flux.
To reduce the noise level in the contours, the image they were taking from was binned by a factor of two. 
The region of strongest [O{\sc iii}] emission is located in the southwest and appears to be composed of two smaller lobes.
If we define the centre of emission as the middle point between these two smaller lobes, then it is located 2.84$\pm$0.14 arcsecs from the position of the binary, with the uncertainty indicating one-pixel uncertainty in the position.

We can compare this value with the previous measurements from T23. From the GMOS LSS data, the extension of the outer shell is 2.33$\pm$0.08 arcsec, while from the GMOS NB image, the extension of the brightest region in the outer shell is 1.91$\pm$0.10 arcsec. Both values are lower compared with our measurement from the MUSE data.
In the case of the LSS measurement, the discrepancy can be explained by the fact that the angle of the used slit did not perfectly coincide with the angle of the outer shell.
In the case of the NB image, the difference occurs because the contours are tracing only the material that is expanding orthogonally to us, while the NB image includes the blue and red components of the ejecta. This causes the brightest region in the NB image to be shifted closer to the binary.

In fact, if we compare the GMOS NB images from T23 with bandpass images from MUSE that mimic the GMOS NB images we can observe there are no major differences between them (Fig.~\ref{fig:gmos_vs_muse}). The figure shows the MUSE NB images for H$\rm\alpha$ and [O{\sc iii}]\,$\rm\lambda$\,5007\,{\AA} in grey, with contour levels indicating the extension of both shells for the MUSE (red) and GMOS (cyan) data. 
The images were normalized in flux, with the total normalized flux in right ascension and declination presented at the top and right of the images.
The levels correspond to 0.1, 0.5, and 0.8 times the maximum normalized flux value in the case of H$\rm\alpha$, and 0.4, 0.6, and 0.8 times in the case of the [O{\sc iii}] images. For the latter, the images were binned by a factor of two before drawing the contour levels to reduce the noise in the data.
The GMOS and MUSE data for the H$\rm\alpha$ images behave identically, with no significative differences. There are minor differences in the case of the [O{\sc iii}] line, particularly in the region of the highest emission which in the GMOS data appears slightly further south than in the MUSE data. We attribute this spatial difference to the lower S/N in the GMOS data and we do not consider it to be significant.
From the MUSE data, we determine the position of the highest emission observed in the [O{\sc iii}] NB image to be located at 2.0$\pm$0.2 arcsec and at an angle of 222$\pm$6 degrees with respect to the centroid of the emission observed in H$\rm\alpha$, in good agreement with the values derived in T23. 

\subsection{Channel maps}

The MUSE capabilities allow us to observe the behaviour of the nebular structure in the inner and outer shell as a function of the velocity of the material.
This can be seen in the channel maps of the lines of H$\rm\alpha$, [N{\sc ii}]\,$\rm\lambda$\,6584\,{\AA}, and [O{\sc iii}]\,$\rm\lambda$\,5007\,{\AA} (top, middle, and bottom row of Fig.~\ref{fig:channel_maps}).
They show the observed emission at a given wavelength after performing a continuum subtraction.
To emphasize the different features observed at each emission line we selected a specific velocity range for each one.
A full wavelength view of the channel maps is presented in the appendix~\ref{sec:appendix_channelmaps}.

The inner shell traced by the H$\rm\alpha$+[N{\sc ii}] emission shows a general symmetric distribution around the binary that also includes several regions of clumpy material that are distributed non-homogeneously across it.
At velocities close to the H$\rm\alpha$ rest wavelength (top row) a ring-like structure can be discerned, which, however, becomes disrupted both at redder and bluer velocities. Instead, regions of higher emission appear within the shell without a clear preferential axis or defined structure.
Additional clumpiness, likely related to [N{\sc ii}]\,$\rm\lambda$\,6584\,{\AA} rather than H$\rm\alpha$, can be observed at longer wavelengths (middle row). Some of these clumps appear to have a preferential axis that coincides with the axis of the outer ejecta (northeast-southwest), although the quality of the data does not allow us to claim this with confidence.

\begin{figure*}[h]
    \centering
    \includegraphics[width=1.0\textwidth]{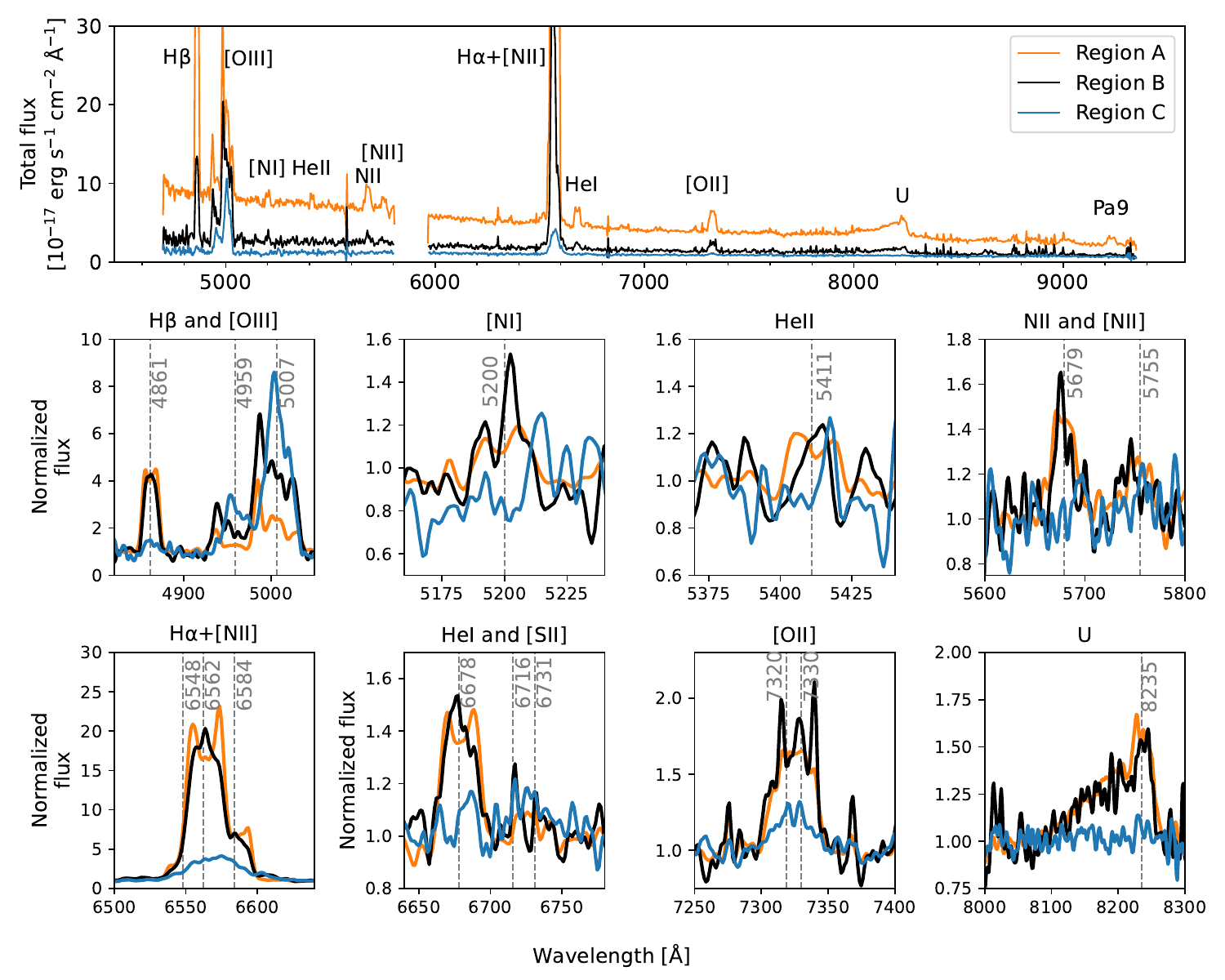}
    \caption{Extracted spectra of the three different regions defined: A (binary + inner shell), B (inner shell only), and C (outer shell).
    The top panel shows the extracted spectrum to illustrate the differences between them along the whole MUSE wavelength coverage in terms of the continuum and emission lines.
    The bottom panels show a zoomed view of the differences between the three spectra for the different lines, including some weak lines that are not visible in the top panel spectra. All the bottom spectra were normalized to the continuum before being presented for a better comparison.
    }
    \label{fig:extracted_spectra}
\end{figure*}

The outer shell is traced mainly by the line of [O{\sc iii}]\,$\rm\lambda$\,5007\,{\AA} (bottom row), but it is also possible to detect a faint tail associated with it in the [N{\sc ii}]\,$\rm\lambda$\,6584\,{\AA} line.
The spatial position of the [O{\sc iii}] emission strongly depends on its velocity, with the bluest and reddest parts being closer to the position of the binary, while its maximum extension occurs when its velocity is closer to zero.
We also note that the flux observed at the position of the binary (white cross) lacks the flux of the red component, with only its blue component present.
This behaviour is not consistent with a cone or plume structure and instead is more consistent with an arc-shaped structure that partially encloses the inner shell.
Regarding the [N{\sc ii}] emission associated with the outer shell, this appears as a blob of material that shows signals of separation with the main emission at the centre. The separation in question appears as a small region with a clear decrease in the observed flux. This confirms the idea that the inner and outer shells are in fact two different components.

\subsection{Shell spectra}

One of the advantages of MUSE is that we can select the spaxels within the datacube to maximize the signal from the different shells when extracting their spectrum, while also minimizing unwanted contributions from other sources. 
We defined three different regions, each one chosen to maximise the contribution of the different components of the system (inner and outer shells, and the binary itself).
We used the contour maps presented in Fig.~\ref{fig:shell1} for the lines of H$\rm\alpha$ and [O{\sc iii}] as a reference to define the extension of these regions.
The spaxels inside each region were summed to create a spectrum that should be representative of the component in question, and they were smoothed by convolving them with a Gaussian profile with a Full-Width at Half Maximum (FWHM) of 3.75\,{\AA} (3 pixels) to reduce the noise.

The first region (region A) corresponds to a spectrum of the binary extracted using an aperture radius of 1.4 arcsec. We obtain this value by modelling the flux in the i\_SDSS bandpass image using a Gaussian function from which we determined a $\sigma=0.48$. The selected aperture corresponds to three times this value, therefore enclosing most of its flux. We must note that because the inner shell is also present in the line of sight of the binary, this will also include features from it. Lastly, we masked the nearby star at the southeast using a mask of 1 arcsec radius to avoid undesired contamination. A total of 136 spaxels were used to extract this spectrum.

The second region (region B) was extracted considering the spaxels that are contained within the contour level corresponding to 0.1 times the maximum flux observed in H$\rm\alpha$ and outside the 1.4 arcsec aperture used in the region A. As for the previous spectrum, we masked the contribution from the nearby star. As the stellar contribution of both the nearby star and the binary has been removed, this spectrum contains only emission from the inner shell but misses the emission from the ejecta in the line of sight. A total of 175 spaxels were selected.

The last region (region C), corresponding to the outer shell, was extracted from the spaxels within the 0.5 contour level observed in [O{\sc iii}] and outside of the 0.1 contour level drawn from the H$\rm\alpha$ emission. The latter criterion was applied to minimize the contamination from the inner shell into the spectrum. In this case, a total of 49 spaxels meet our requirements.

\begin{table*}[h]
    \centering
    \caption{Fluxes and S/N for the different lines in the extracted spectra from the different regions considered, as well as the images corresponding to each line. All fluxes are given in units of erg\,s$^{-1}$\,cm$^{-2}$ with the values in parenthesis indicating the uncertainty in the last significant digit.}
    \resizebox{1.0\textwidth}{!}{
    \label{tab:fluxes}
    \begin{tabular}{cccccccccc}
        \hline\hline\noalign{\smallskip}
        Line&Rest wavelength&\multicolumn{2}{c}{Region A}&\multicolumn{2}{c}{Region B}&\multicolumn{2}{c}{Region C}&\multicolumn{2}{c}{Image}\\
        &[\AA]&Flux&S/N&Flux&S/N&Flux&S/N&Flux&S/N\\
        \hline \noalign{\smallskip}
        H$\rm\beta$&4861&4.79(4)$\times$10$^{-15}$&128&1.55(3)$\times$10$^{-15}$&48&--&--&7.5(2)$\times$10$^{-15}$&39\\
        
        [O{\sc iii}]&4959,5007&5.82(6)$\times$10$^{-15}$&102&5.94(5)$\times$10$^{-15}$&119&2.67(3)$\times$10$^{-15}$&96&2.10(3)$\times$10$^{-14}$&78\\
        
        [N{\sc i}]&5197,5200&9(2)$\times$10$^{-17}$&5&--&--&--&--&--&--\\
        He{\sc ii}&5411&1.2(2)$\times$10$^{-16}$&6&--&--&--&--&--&--\\
        N{\sc ii}&5679&7.0(4)$\times$10$^{-16}$&18&2.5(3)$\times$10$^{-16}$&8&--&--&1.5(1)$\times$10$^{-15}$&14\\
        
        [N{\sc ii}]&5755&3.4(4)$\times$10$^{-16}$&9&1.7(3)$\times$10$^{-16}$&5&--&--&1.2(1)$\times$10$^{-15}$&11\\
        H$\rm\alpha$+[N{\sc ii}]&6562+6548,6584&2.690(3)$\times$10$^{-14}$&961&8.22(2)$\times$10$^{-15}$&369&1.118(8)$\times$10$^{-15}$&140&4.65(9)$\times$10$^{-14}$&495\\
        He{\sc i}&6678&4.7(2)$\times$10$^{-16}$&21&1.8(2)$\times$10$^{-16}$&10&--&--&9.7(6)$\times$10$^{-15}$&15\\
        
        [S{\sc ii}]&6716,6731&--&--&--&--&2.7(5)$\times$10$^{-17}$&6&--&--\\
        
        [O{\sc ii}]&7320,7330&7.7(1)$\times$10$^{-16}$&57&3.2(1)$\times$10$^{-16}$&28&5.2(5)$\times$10$^{-17}$&11&1.49(5)$\times$10$^{-15}$&28\\
        U&$\sim$8235&2.60(4)$\times$10$^{-15}$&66&7.7(4)$\times$10$^{-16}$&21&--&--&3.8(1)$\times$10$^{-15}$&24\\
        Pa9&9229&3.0(2)$\times$10$^{-16}$&15&--&--&--&--&--&--\\
        \hline
    \end{tabular}
    }
\end{table*}

The extracted spectra can be observed in Fig.~\ref{fig:extracted_spectra}, where the top panel shows the spectra across the full MUSE wavelength range, and the bottom panels the differences between spectra for some emission lines.
The spectrum corresponding to the region A (orange) shows a bluer continuum compared with the other two spectra, which must come from the binary. It also shows several lines, the most prominent ones being the Balmer (H$\rm\beta$ and H$\rm\alpha$) and [O{\sc iii}]\,$\rm\lambda\rm\lambda$\,4959,5007\,{\AA} lines, but also contributions from He{\sc i}\,$\rm\lambda$\,6678\,{\AA} and He{\sc ii}\,$\rm\lambda$\,5411\,{\AA}, as well as nitrogen lines: [N{\sc i}]\,$\rm\lambda\rm\lambda$\,5197,5200\,{\AA}, [N{\sc ii}]\,$\rm\lambda$\,5755\,{\AA}, [N{\sc ii}]\,$\rm\lambda\rm\lambda$\,6548,6584\,{\AA}, and N{\sc ii}\,$\rm\lambda$\,5679\,{\AA}.
Compared with the spectra presented in T23, the MUSE data reach redder wavelengths which allows us to detect the lines of [O{\sc ii}]\,$\rm\lambda\rm\lambda$\,7320,7330\,{\AA} and Paschen9 at 9229\,{\AA}. We also observe an emission feature at $\rm\lambda\sim$\,8235\,{\AA} which we were not able to identify. Its profile shape suggests a blend with possible candidates being the lines of He{\sc ii}\,$\rm\lambda$\,8237\,{\AA}, N{\sc i}\,$\rm\lambda$\,8216,8223 or 8242\,{\AA}, and [O{\sc i}]\,$\rm\lambda$\,8222\,{\AA}. From here on, we will refer to this emission as the U (for unidentified) emission.

Because the region A spectrum includes both the binary and the inner shell, all the observed forbidden emission in this spectrum must originate in the latter. 
This is supported by the spectrum of region B, which corresponds to the inner shell only, showing all the forbidden lines observed in the spectrum of region A. Allowed transitions such as Balmer or He{\sc i}, whose presence is not rare in nova shells, are also present in this spectrum. The He{\sc ii} and Pa9 lines observed in the spectrum of region A are not observed here, which indicates that these two lines originated in the binary, likely from an accretion disk around the WD, and not from the inner shell. The unidentified emission is also present in this spectrum, implying its origin in the inner shell.

The outer shell from region C is dominated by the [O{\sc iii}] emission instead of H$\rm\alpha$+[N{\sc ii}] as is the case in the other two regions. As was described by T23, there is no significant contribution of Balmer lines in the outer shell, and this is also valid for the He and N lines with the exception of [N{\sc ii}]\,$\rm\lambda\rm\lambda$\,6548,6584\,{\AA}.
The newly observed line of [O{\sc ii}] shows a minor contribution only in this spectrum.
Lastly, the spectrum of the inner shell (black) shows a weak continuum which is likely the residual from the masked nearby star, while the outer shell spectrum (blue) does not show any noticeable continuum contribution.

The bottom panels of Fig.~\ref{fig:extracted_spectra} show a closer look at several lines. 
For a better visualization of the differences in the spectral lines, they were normalized to the continuum.
Although the spectra of regions A and B share many similarities, the inclusion of the outermost part of the inner shell in the later spectra causes differences in the shape of the different lines.
This is noticeable in all lines, but particularly evident in the lines of H$\rm\alpha$ and He{\sc i}\,$\rm\lambda$\,6678\,{\AA}. In these cases, the spectrum corresponding to region A shows distinctive double peak profiles originating from the expanding shell. When we consider only the material from the outer part of the inner shell instead (region B), the double peak is replaced with a single peak emission.
The normalized spectra show similarities between the lines of N and He, with both of them being absent in the outer shell spectra and showing stronger normalized fluxes in the spectrum of region B, indicating that the two lines originate in the inner shell.

Right next to the He{\sc i} line, we can observe a weak emission that could tentatively be identified as [S{\sc ii}]\,$\rm\lambda\rm\lambda$\,6716,6731\,{\AA} doublet, although it must be noted that the intensity at these wavelengths only barely exceeds the continuum noise, slightly more so in region C than in the other two.
The profile observed in the [O{\sc ii}]\,$\rm\lambda\rm\lambda$\,7320,7330\,{\AA} is present in the three spectra but shows significant differences among them. It appears stronger in the inner shell spectra where three distinct peaks are present, only the middle one coinciding with the rest wavelength. However, the continuum in this spectrum also shows similar peaks that we attribute to noise in the data, which sheds some doubt on the significance of the above structure in the emission line. In the case of region A, the line profile shows a single peak profile, while in the case of the outer shell, two small peaks at the rest wavelength can be observed.
Lastly, the emission at $\rm\lambda\sim$\,8235\,{\AA} is present only in the binary and inner shell spectra with very similar profiles in both cases, which suggests that its origin can be found in the inner shell rather than in the binary.

\begin{figure}[h]
    \centering
    \includegraphics[width=0.95\columnwidth]{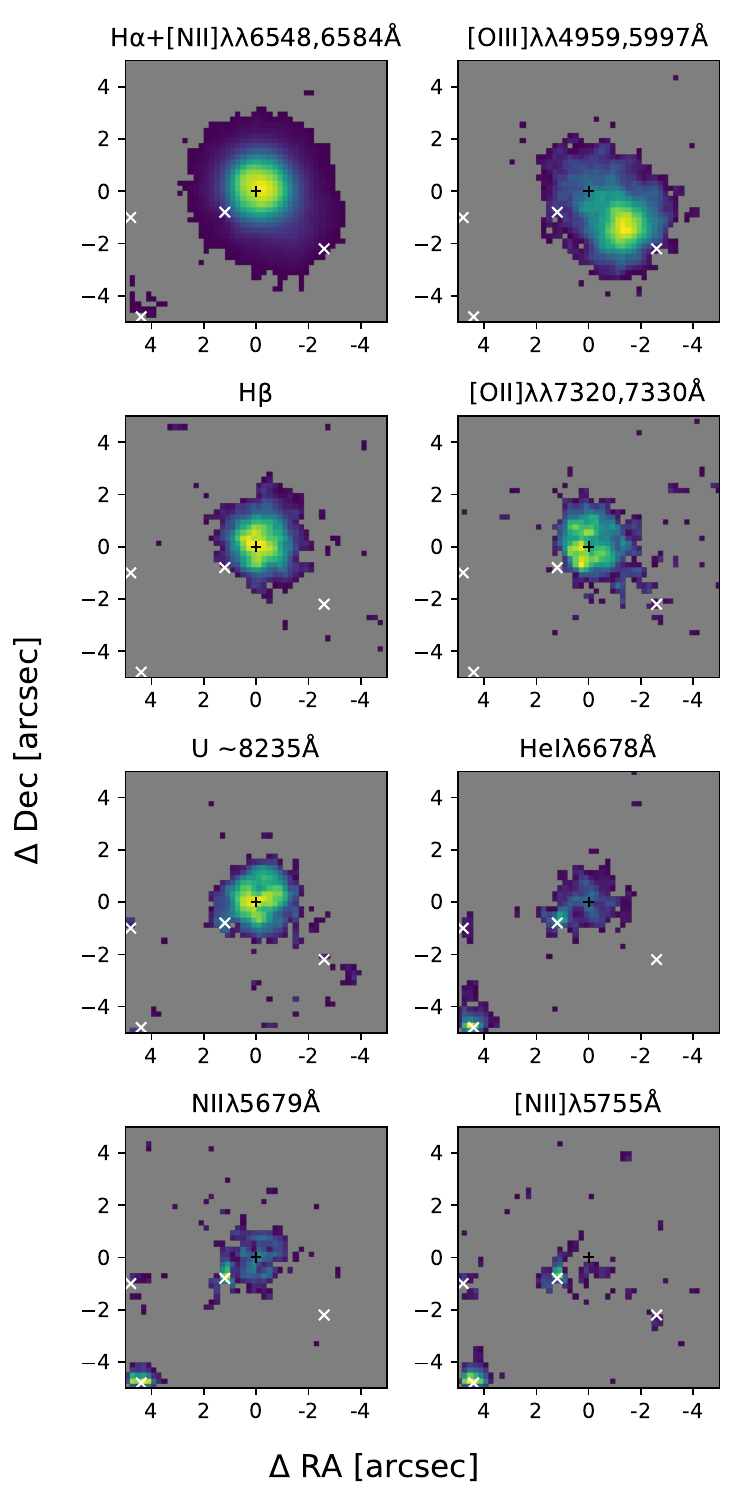}
    \caption{MUSE images for the strongest lines in the V1425 Aql nova shell. They were obtained by summing the flux within a certain spectral range after subtracting the continuum. They were smoothed using a Gaussian kernel and the values below the background median value plus one sigma were masked. The position of the binary is marked with a black cross, while the white marks indicate the position of the nearby stars in the field. Most of the observed emission traces the inner shell, while the outer one is observed only in [O{\sc iii}], H$\rm\alpha$+[N{\sc ii}], and [O{\sc ii}].}
    \label{fig:mean_lines}
\end{figure}

\subsection{Fluxes} \label{sec:fluxes}

The fluxes of the most prominent lines in the three spectra are presented in Table~\ref{tab:fluxes}. They were obtained by integrating the spectra within a velocity range of $\pm$1500 km s$^{-1}$ after performing a subtraction of the continuum in the form of a one-degree polynomial. 
In the case of the lines of [O{\sc ii}]\,$\rm\lambda\rm\lambda$\,4959,5007\,{\AA} and H$\rm\alpha$+[N{\sc ii}]\,$\rm\lambda\rm\lambda$\,6549,6584\,{\AA} the velocity range was increased to $\pm$2000 km\,s$^{-1}$ and the respective lines combined to incorporate all the flux. On the other hand, for the line of [S{\sc ii}]\,$\rm\lambda$\,6716,6731\,{\AA}, we reduce the spectral range up to $\pm$500 km\,s$^{-1}$ to avoid contamination from the nearby He{\sc i}\,$\rm\lambda$\,6678\,{\AA} line. For the U emission, we consider a spectral range between 8000 and 8300\,{\AA}, as we were not able to identify the origin of this emission.
Furthermore, we also measured the total incoming flux at each line, without distinguishing between spectra.
For that purpose, we obtain an image from the datacube corresponding to the same spectral range used to determine the line fluxes. The image was obtained by subtracting the continuum and then summing the fluxes within a circular aperture of 5 arcsec.
The results are also presented in Table~\ref{tab:fluxes}.
Not all the lines are present in all the spectra, so we present only the fluxes of the lines that have a $S/N \geq 5$ in the case of the spectra and $\geq10$ for the images. 

The extension of the shell in the different lines is presented in Fig.~\ref{fig:mean_lines}.
Each presented image was smoothed by convolving them with a 2D symmetric Gaussian kernel of FWHM of 0.2 arcsecs (1-pixel size), and pixels whose flux is below the median level of the background plus its standard deviation were discarded.
In all cases, a linear scale was used which, in combination with the exclusion of the faintest pixels, provides a good visualization of the nova shell.
The position of the binary is marked with a black cross, while the positions of the closest stars to V1425 Aql are indicated with white x marks.

The images are sorted from the highest (top left panel) to the lowest (bottom right panel) flux. 
As had been established in T23, the strongest lines are H$\rm\alpha$+[N{\sc ii}] and [O{\sc iii}], followed by the H$\rm\beta$ line.
The MUSE data unveiled the presence of the lines of [O{\sc ii}]\,$\rm\lambda\rm\lambda$\,7320,7330\,{\AA} and the U emission which also present a significant flux. Both emission lines are observed tracing mainly the inner shell, with the [O{\sc ii}] emission also showing a small contribution to the outer shell.
Next in flux are the previously detected allowed transitions of He{\sc i}\,$\rm\lambda$\,6678\,{\AA} and N{\sc ii}\,$\rm\lambda$\,5679\,{\AA}, which are observed to trace the inner shell only.
The last line presented is [N{\sc ii}]\,$\rm\lambda$\,5755\,{\AA} showing only a spurious detection of the inner shell.
This line also illustrates how the contribution of the nearby stars starts to dominate as the flux of the emission line becomes fainter. For lines weaker than [N{\sc ii}] the images do not show any emission that can be related to the shell, inner or outer.

\subsection{Shell geometry}

The geometry of the nova shell around V1425 Aql appears to be complex, with the inner and outer shells showing evident differences in their geometries. The channel maps presented in Fig.~\ref{fig:channel_maps} also indicate the presence of structures hidden within the inner shell.
To discern the geometry of the shell we can study the Position-Position-Velocity (PPV) space, and in particular the Position-Velocity (PV) planes. The PPV space corresponds to the three axes in the datacube: the equatorial coordinates (right ascension and declination), together with the spectral range. If we collapse the PPV space into one of their equatorial coordinates we end up with an image of one of the datacube's faces: a PV plane.
The PPV space and the PV planes allow us to visualize the spatial and spectral extension of the nova shell.

Additionally, the PPV can be converted into a proper physical space if we know the distance to the system, the time since the nova eruption, and its systemic velocity. We refer to this physical space as Position-Position-Position (PPP) space. In \citet{Celedon+2024A&A...681A.106C} we demonstrate that under the assumption of free and radial expansion, the conversion from the PPV to the PPP is a linear function. This allows us to determine the physical extension of the inner and outer shells.

\subsubsection{PV planes}

\begin{figure*}
    \centering
    \includegraphics[width=0.95\textwidth]{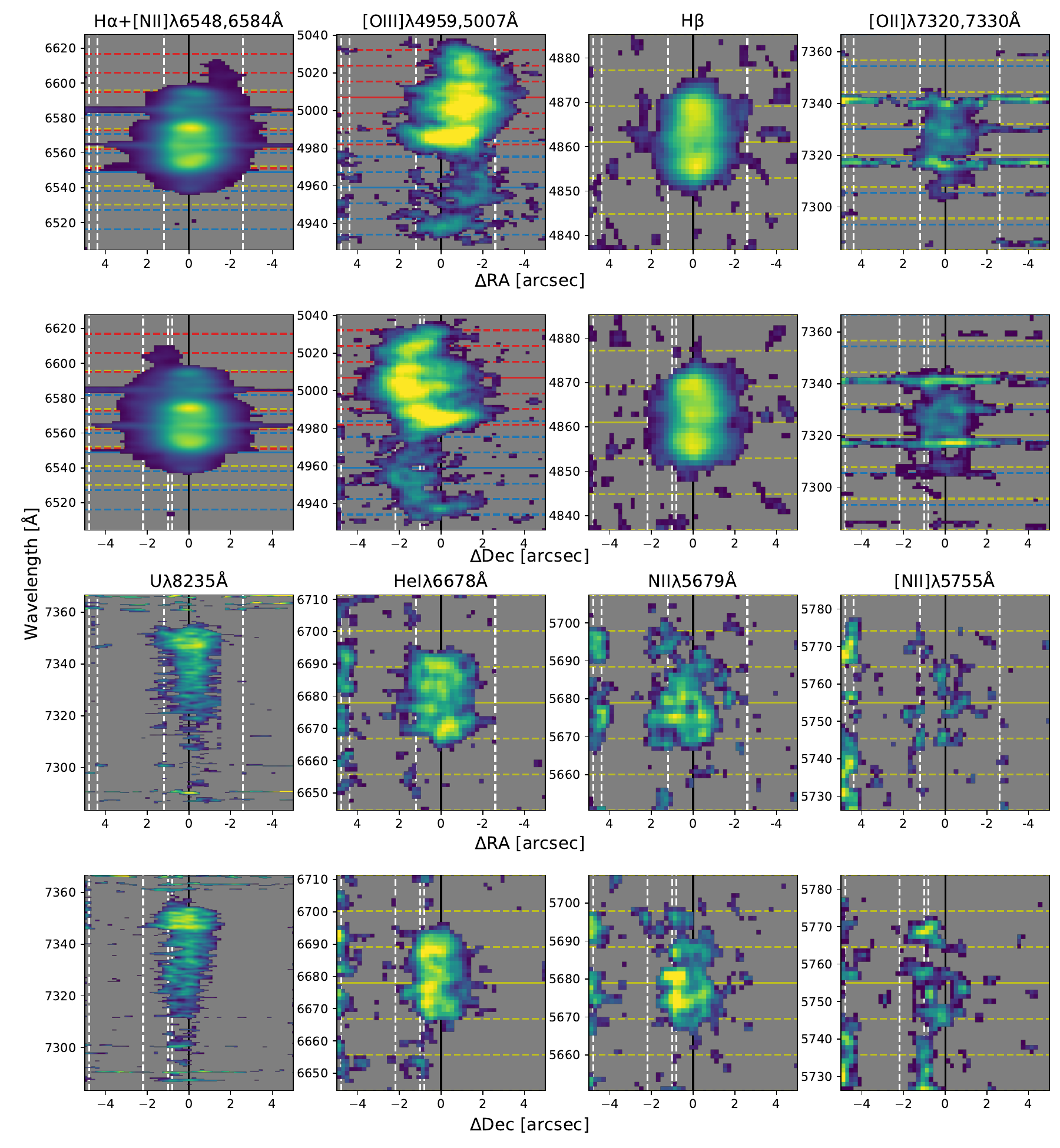}
    \caption{PV planes for RA and Dec showing the fluxes of H$\rm\alpha$+[N{\sc ii}], [O{\sc iii}],  H$\rm\beta$, the U emission, [O{\sc ii}], N{\sc ii}, He{\sc i}, and [N{\sc ii}]. In each plane, the horizontal dashed lines indicate velocities in steps of $\pm$500 km\,s$^{-1}$ with respect to the rest line (solid line), while the positions of the nearby stars to V1425 Aql are marked with vertical white dashed lines.
    In the RA planes east is up to the left, while in the Dec planes north is up to the right.
    To highlight the flux from the nova shell the planes were smoothed and the pixels with low fluxes were discarded. In all cases, the fluxes are presented using a linear scale.
    All lines observed trace the inner shell, which appears as a compact, symmetric, and slow ($v_{\rm{exp}}\lesssim$500 km\,s$^{-1}$) component.
    The outer shell appears clearly in the [O{\sc iii}] planes as an asymmetric component, but also in [N{\sc ii}] as a little horn in the red part of the H$\alpha$+[N{\sc ii}] planes.
    }
    \label{fig:pv_planes}
\end{figure*}

We extract the expanding shell from the datacube (that is, the PPV space) by following a similar procedure as we did for the images in Sect.~\ref{sec:fluxes}.
We consider the same spectral ranges as we previously did, and we subtracted the continuum in the same way (by fitting a 1-degree polynomial).
Once the shell has been extracted, we can proceed to extract the PV planes corresponding to the RA and Dec.
They were obtained by summing the PPV in the Dec and RA axes respectively.
To clean the PV planes and highlight the different structures within the shell, we smooth the resulting planes by convolving them with a Gaussian kernel of FWHM equal to 2 pixels of the image. We discarded points in the plane whose fluxes are below the median value plus one standard deviation of the background.
We applied this extraction to the same lines as in Fig.~\ref{fig:mean_lines}.
The resulting PV planes for right ascension and declination are presented in Fig.~\ref{fig:pv_planes}.

Each PV plane shows the observed flux using a square-root scale, with the yellow dashed lines indicating velocities of $\pm$500 km\,s$^{-1}$ with respect to the rest wavelength.
In case two or more lines are involved, they are represented by blue and red colours (two lines involved) or by blue, yellow, and red colours (three lines involved), from shorter to longer wavelengths.
For each line, both planes are presented, the right ascension (upper row) and declination (lower row) planes. The position of the nearby stars is marked with white dashed lines, while the binary lies at position zero (black line).

The flux distribution observed in each plane for the different lines allows us to draw conclusions regarding its geometry.
Starting with the strongest line, H$\rm\alpha$ shows a compact source with two prominent lobes at red and blue wavelengths. This structure can be better appreciated in the H$\rm\beta$, where even a small ring structure starts to appear in the centre of the shell. 
The [N{\sc ii}]\,$\rm\lambda$\,6584\,{\AA} emission at its rest wavelength appears to be stronger at the north and east while it starts to appear closer to the binary position at redder and bluer wavelengths, in concordance with what is observed in the channel maps.
This behaviour is similar to what is observed in the [O{\sc iii}] emission tracing the outer shell, although in this case the emission at rest wavelength is observed at the southwest. It should be noted that the behaviour described for the [N{\sc ii}] emission is observed within the inner shell, which may indicate a common ejection mechanism for both emissions.
The inner shell can be observed in [O{\sc iii}]\,$\rm\lambda$\,5007\,{\AA} as the central structure enclosed within $\pm500$km\,s$^{-1}$. An asymmetric flux is evident in the inner shell, better seen in the Dec plane, where the blue part is stronger than its red counterpart. This is also noticeable in the outer shell, which appears as a 'banana' shape in the [O{\sc iii}] planes. The [N{\sc ii}] also reveals a hint of the outer shell in the form of a little horn at the reddest wavelengths of the plane.

The rest of the planes are noisier, and the shell is not evident in all of them. The plane for the [O{\sc ii}]\,$\rm\lambda\rm\lambda$\,7320,7330\,{\AA} lines show two evident artefacts product of wavelengths with high noise across the entire datacube. The inner shell can be observed as a compact source in the planes, with the blue component of the line at $\rm\lambda$\,7320\,{\AA}, the central region where both lines contribute, and the red part of the $\rm\lambda$\,7320\,{\AA} being distinguished. The emission related to the outer shell is not observed. This shows that the PV planes are less sensitive to the faintest part of the emission when compared with spectra or images from the datacube.
The U emission is clearly observed, and in the same manner as the spectra showed, it starts to increase its flux from blue to redder wavelength until reaches its peak at $\sim$8225\,{\AA}. The feature does not show any remarkable asymmetry in the planes.

The inner shell can still be distinguished within the PV planes of the He{\sc i}\,$\rm\lambda$\,6678\,{\AA} and N{\sc ii}\,$\rm\lambda$\,5679\,{\AA} lines, but they also show a significant contribution in flux from the nearby stars.
In both cases, the compact structure associated with the inner shell can still be recognized despite the stronger contamination from nearby stars whose positions in the planes are marked with white dashed lines. These stars are located mainly at the south (to the left in the Dec planes) and east (to the left in the RA planes) of the binary, leaving the north and west of the planes largely uncontaminated. The emission observed in these regions, thus, must come mainly from the inner shell.
This is relevant for the plane of [N{\sc ii}]\,$\rm\lambda$\,5755\,{\AA} which is the one which shows the large contamination from nearby stars. Regardless, a faint emission is observed to be located close to the position of the binary, including some emission to the north and west.

\subsubsection{Position-Position-Position}

We applied this conversion to the extracted shell observed in [O{\sc iii}]\,$\rm\lambda$\,5007\,{\AA} and H$\rm\beta$. We chose these lines because each one traces one of the shells (H$\rm\beta$ the inner one and [O{\sc iii}] the outer shell), they both have a good S/N, and they do not blend with other lines.
The latter is of particular importance, as for blended lines the line identification (and therefore the conversion to velocity) is not trivial, which could cause artefacts in the transformation to spatial position.
For the transformation from PPV to PPP, we also need the distance and systemic velocity of the binary. We used the values determined in T23 corresponding to 3.3 kpc and 45 km\,s$^{-1}$ respectively.

\begin{figure}
\centering
\includegraphics[width=0.95\columnwidth]{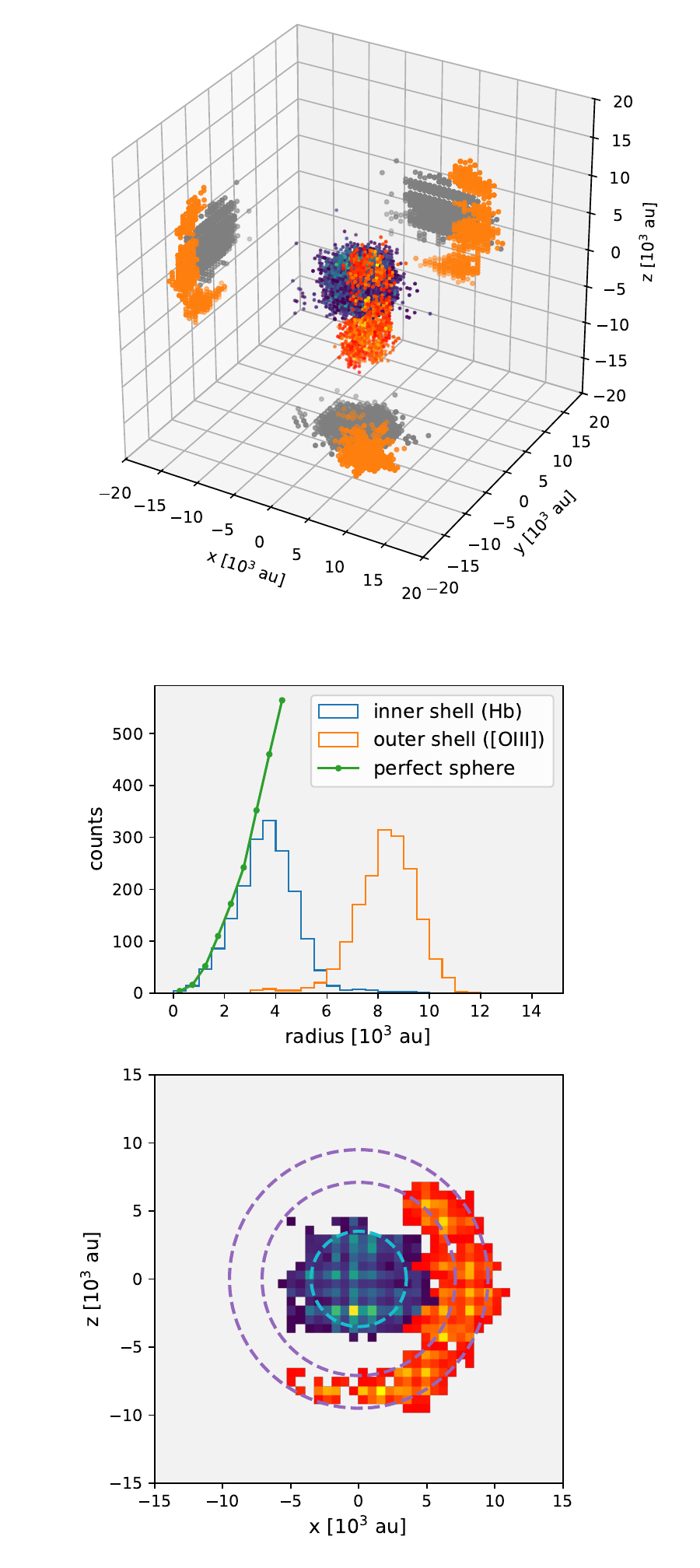}
\caption{
Three-dimensional reconstruction of the outer shell (red) traced by the line of [O{\sc iii}] at 5007\,{\AA}, and the inner shell (blue) traced by the H$\beta$ emission.
The top panel shows the projection of the inner and outer shell in the PPP space, together with their respective projections in each one of the planes, in orange and grey respectively.
The histogram of the radial distance of both shells from the position of the central binary is presented in the middle panel.
The bottom panel shows a cut of both shells at the angle of the outer shell. The data was rotated and interpolated accordingly to present the structure of both shells.
The data indicate a spherical inner shell that extends up to $\sim3\,500$\,AU (cyan dashed line), while the outer shell shows an arc-like geometry that extends from $\sim7\,100$ up to $\sim9\,500$\,AU (purple dashed lines).
}
\label{fig:outer_shell_3d}
\end{figure}

The resulting PPP is shown in Fig.~\ref{fig:outer_shell_3d}. 
A projection of the 3D reconstruction is presented in the top panel of the figure, where the different shells are presented in different colourmaps: red for [O{\sc iii}] and blue for H$\rm\beta$.
The plot also includes the projection of the shell in the different planes.
The 3D reconstruction of the outer shell indicates that it possesses an arc-shaped geometry that partially encircles the inner shell.
This is consistent with what is observed in the channel maps (Fig.~\ref{fig:channel_maps}) and the PV space (Fig.~\ref{fig:pv_planes}) of the line of [O{\sc iii}].

On the other hand, the H$\rm\beta$ emission presents a geometry that is closer to a sphere. 
This is supported by the behaviour observed in the histogram of the radial distances of the shell (middle panel), which was obtained after transforming the cartesian coordinates of the PPP into spherical coordinates.
To compare the observed distribution of the inner shell (blue) in the histogram, we recreated the behaviour that a perfect solid sphere (green) should have, that is, the number of counts as a function of the radius given the same spatial distribution observed in the PPP.
The comparison shows that the inner shell closely follows the distribution of a perfect sphere for radii lower than $3\,500$\,AU. Beyond this radius, the distribution starts to deviate significantly from a sphere, thus, defining its outer limits.
The outer shell shows a distribution close to a Gaussian slightly skewed to lower radii, with the most likely reason being the contribution of the inner shell observed in [O{\sc iii}]. 
The mean ($\mu$) and standard deviation ($\sigma$) of this distribution are $\sim8\,300$ and $1\,200$\,AU respectively.

The last panel in the figure presents a cross-section of 1 arcsec width of the expanding shell at the angle of the outer shell (222 degrees). The data was rotated to present a front view of this cross-section, with the PPP's z-axis (line-of-sight axis) presented in the y-axis of the plot while the x-axis in the plot is a mixture of the PPP's x and y-axis.
The view of the shell from this angle confirms once more the arc-like structure of the outer shell, and how it spans an aperture angle of $\sim$180 degrees. 
The radius of $3\,500$\,AU determined for the inner shell is presented with a cyan dotted line. It encloses the inner shell well on the z-axis but not so well on the x-axis. This may be an indication of an overestimation of the distance used to obtain the PPP space.
Similarly, the distribution of the [O{\sc iii}] emission shows a bulk in the x-axis of the plot close to z$\sim$0, but in this case, we could attribute it to the presence of the Oxygen inner shell.
In general, the outer shell is well confined within the radius defined by $\mu\pm\sigma$ (purple dashed lines). 

\section{Discussion}

The main result of this work is the finding that the geometry of the outer shell corresponds to an arc-shaped structure that partially encircles the inner, and more spherical, shell.
This is unexpected, as it was presumed that the most likely geometry for an isolated ejection should be a cone or a plume.
A possible physical explanation for both the observed asymmetry and the arc-shaped geometry of the outer shell could be found in a magnetic WD.
The presence of a magnetic WD in V1425 Aql was proposed by \citet{Retter+1998MNRAS.293..145R} based on a photometric analysis of the system after the nova eruption. They found strong periodic signals at $\sim$6.1 and $\sim$1.4 h, allegedly corresponding to the orbital period and the WD spin respectively. However, \citet{Worpel+2020A&A...639A..17W} did not detect any X-ray emission, thus challenging the previous idea. Therefore, the magnetic nature of the WD in V1425 Aql is not settled. Still, by assuming it is, we can speculate how the observed asymmetry could find its origin in a magnetic WD.

In magnetic CVs where the magnetic and orbital axes of the WD are misaligned, the accretion rate on the magnetic poles may differ between them \citep{Zhilkin+2022Galax..10..110Z}. This will cause one of the poles to be hotter after the nova eruption, causing the thick winds that produce the fast outflow to be stronger at one of the poles.
If the thick wind phase lasts for only half of the WD spin, this could in principle also explain the arc-like structure observed in the outer shell.
Lastly, another piece of circumstantial evidence in favour of a magnetic WD is given by the spherical geometry of the inner shell. \citet{Porter+1998MNRAS.296..943P} investigated the effect of the WD rotation in the shape of the nova shell, finding that slower rotating WDs produce more spherical shells. Assuming that the period of 1.44 hours modulation detected by \citet{Retter+1998MNRAS.293..145R} corresponds to the spin period of the WD, this would classify the WD in V1425 Aql as a very slow rotator \citep{Norton+2004ASPC..315..216N}.

This idea, however, is not exempt from challenges. While an asymmetric accretion is possible under the circumstances discussed, one as strong as the one observed in the outer shell would require the WD to be a synchronous polar, so all the matter will be accreted on one pole only. This view is incompatible with the results of \citet{Retter+1998MNRAS.293..145R} and \citet{Worpel+2020A&A...639A..17W}.
Furthermore, if the physical reason behind the asymmetry ejecta lies in a magnetic WD, one should expect to observe similar features around other magnetic novae. This, however, is not the case \citep[some famous magnetic novae are DQ Her and GK Per,][]{Santamaria+2020ApJ...892...60S, Harvey+2016A&A...595A..64H}.
However, we point out that the vast majority of nova shells have been observed using H$\rm\alpha$ narrow-band filters only. Under these filters, the nova shell around V1425 Aql would appear as normal as any other shell. It is then plausible that some unexpected features have gone unnoticed in other magnetic novae.
The H$\rm\alpha$ and [O{\sc iii}] nova shell survey carried out by \citet{Downes&Duerbeck2000AJ....120.2007D} includes two, although doubtful, intermediate polars: CP Pup \citep{Mason+2013MNRAS.436..212M} and V842 Cen \citep{Woudt+2009MNRAS.395.2177W, Luna+2012MNRAS.423L..75L, Sion+2013ApJ...772..116S}. Neither of these systems showed ejecta with obvious asymmetric features.

On the other hand, a comparison between the GMOS and MUSE NB images could have the potential to clarify the time of the ejection that gave rise to the outer shell, thus providing hints about its origin. Unfortunately, due to the low S/N of the GMOS images, an evolution of the shell, and consequentially its expansion rate in the plane of the sky, cannot be determined with any confidence.
From the MUSE data, we derived a maximum extension of the outer shell of 2.84$\pm$0.14 arcsec, which is higher than the value reported in T23 from the NB image (1.91$\pm$0.1 arcsec) and LSS (2.32$\pm$0.11 arcsec). We attribute the differences to the wider transmission of the GMOS NB image and the angle of the used slit.
From the measurements of T23, there is a margin of $\pm$2 years between the ejection of the inner and outer shell. With the new measurement, and using the same expansion velocities for the outer ejecta measured in T23 (1500$\pm$20 km\,s$^{-1}$), this value increases up to 4$\pm$3 years, still compatible with a simultaneous ejection of the shells within two sigmas.

The MUSE spectral range allows us to study lines at redder wavelengths compared to T23, adding the lines of [O{\sc ii}]\,$\rm\lambda\rm\lambda$\,7320,7330\,{\AA} and the feature observed at $\rm\lambda\sim$8225\,{\AA}. 
The [O{\sc ii}] lines were previously measured by \citet{Lyke+2001AJ....122.3305L} (L01 from here on). About 2.5 years after the nova eruption, these lines were significantly weaker compared with the rest of the lines in their spectrum like [O{\sc iii}]\,$\rm\lambda$\,5007\,{\AA}. From the fluxes they reported we can determine the ratio [O{\sc iii}]\,$\rm\lambda\rm\lambda$\,4959,5007\,{\AA} / [O{\sc ii}]\,$\rm\lambda\rm\lambda$\,7320,7330\,{\AA} to be $\sim$180. In contrast, our data indicates a much lower value for the same lines, corresponding to $\sim$14 from the flux obtained from the images.
The ratio [O{\sc iii}]/[O{\sc ii}] is sensitive to the ionization parameter and the ionization field \citep[e.g.][]{Kewley+2006MNRAS.372..961K, Nakajima+2014MNRAS.442..900N}, and therefore, the drastic reduction in ratio between the L01 observations and ours indicates a significant decrease in the overall radiation field. This is not surprising as the major source of ionizing photons should be the WD, which will cool with time.

We were not able to identify the emission observed at $\sim$8235\,{\AA}. Possible candidates for this emission are the lines corresponding to He{\sc ii}\,$\rm\lambda$\,8237\,{\AA}, N{\sc i}\,$\rm\lambda$\,8216,8223 or 8242\,{\AA}, and [O{\sc i}]\,$\rm\lambda$\,8222\,{\AA}. 
From the spectra (Fig.~\ref{fig:extracted_spectra}) and images (Fig.~\ref{fig:mean_lines}), it appears clear that the observed emission originates in the inner shell, which implies that the Nitrogen and Oxygen lines are plausible candidates. The He{\sc ii} is a less likely candidate as the inner shell does not show any emission from the He{\sc ii}\,$\rm\lambda$\,5411\,{\AA} observed in the spectrum of the region A.
The shape of the observed unidentified emission suggests a blend of lines and therefore several of the mentioned lines could be contributors.
It is also interesting to note that in the spectrum published by L01, there is no clear emission around this spectral region, although the line of He{\sc ii} was detected. Similar to the case of the [O{\sc ii}] lines, this illustrates the evolution of the nova shell with time.

\section{Summary}

Our analysis of the nova shell around V1425 Aql clarified the geometry of the outer and asymmetric ejecta, revealing an unexpected arc-like structure instead of the presumed cone or plume geometry. 
This unusual geometry is revealed by the data presented in the channel maps, the PV planes, and the 3D reconstruction.
The physical reason behind is still unclear, and we have speculated how a magnetic WD may be the reason behind it. However, the ambiguous evidence regarding the presence of a magnetic WD in V1425 Aql prevents us from conclusively identifying this as the true mechanism.

By selecting different spaxels in the datacube, we were able to better characterize the differences between the inner and outer shell, and the binary itself. The outer ejecta is only visible in forbidden lines ([O{\sc iii}], [N{\sc ii}], and [O{\sc ii}]), while the inner shell shows a mix between allowed and forbidden transitions. An interesting emission, likely a blend, appears in the red part of the spectrum, near 8235\,{\AA}, which we were not able to identify, but at least to constrain its origin to the inner shell.

On the other hand, the strongest lines observed in the inner shell, the blend between H$\rm\alpha$ and [N{\sc ii}], start to reveal clumpy structures within the shell. These clumps are observed closer to the [N{\sc ii}] rest wavelength and may appear to share the same axes in the sky as the outer ejecta, although the current data can not discern if this is the true case. As the shell expands, these clumps should be more evident, thus future observations should be able to discern if the clumps share a common axis with the outer ejecta or not.

Overall, while our study provides important progress with respect to the first report on the outer ejecta by \citet{Tappert+2023A&A...679A..40T} it raises additional questions, regarding its shape, the physical reason behind, and the unidentified emission blend near 8235\,{\AA}. This should provide sufficient motivation for further detailed studies on this interesting nova and its peculiar shells.

\begin{acknowledgements}

The raw and reduced MUSE datacubes used in this work can be freely accessed from the ESO archive.
LC acknowledges economic support from ANID-Subdireccion de capital humano/doctorado nacional/2022-21220607.

\end{acknowledgements}

\bibliographystyle{aa} 
\bibliography{bibliography.bib} 

\onecolumn
\begin{appendix}
\section{Channel maps} \label{sec:appendix_channelmaps}

\begin{figure}[!h]
\centering
\includegraphics[width=0.95\textwidth]{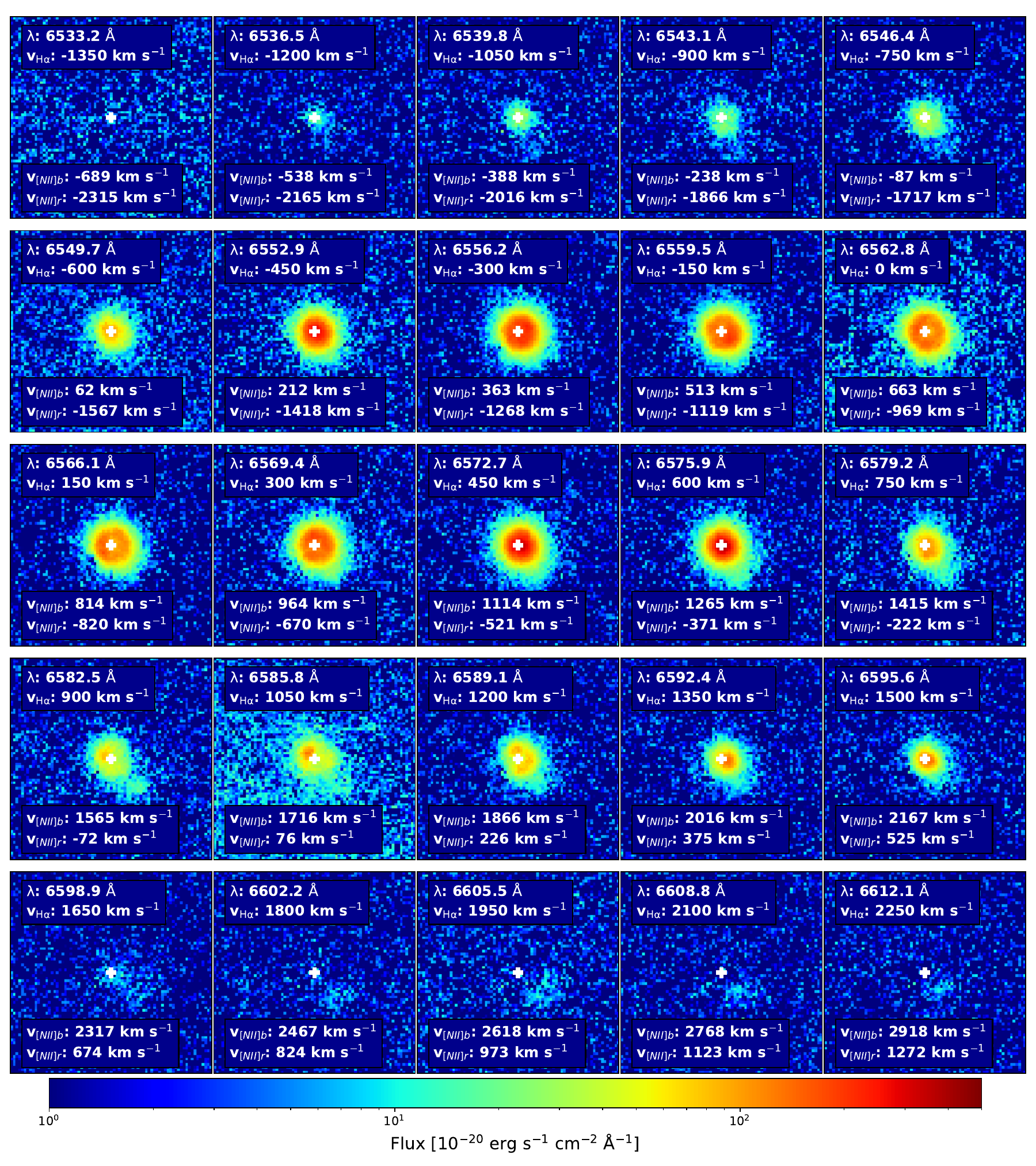}
\caption{Channel maps for H$\rm\alpha$ and [N{\sc ii}]\,$\rm\lambda$\,6548\,{\AA} ([N{\sc ii}]b) and $\rm\lambda$\,6584\,{\AA} ([N{\sc ii}]r) emission lines.}
\label{fig:appendix_1}
\end{figure}

\begin{figure}[!h]
\centering
\includegraphics[width=0.95\textwidth]{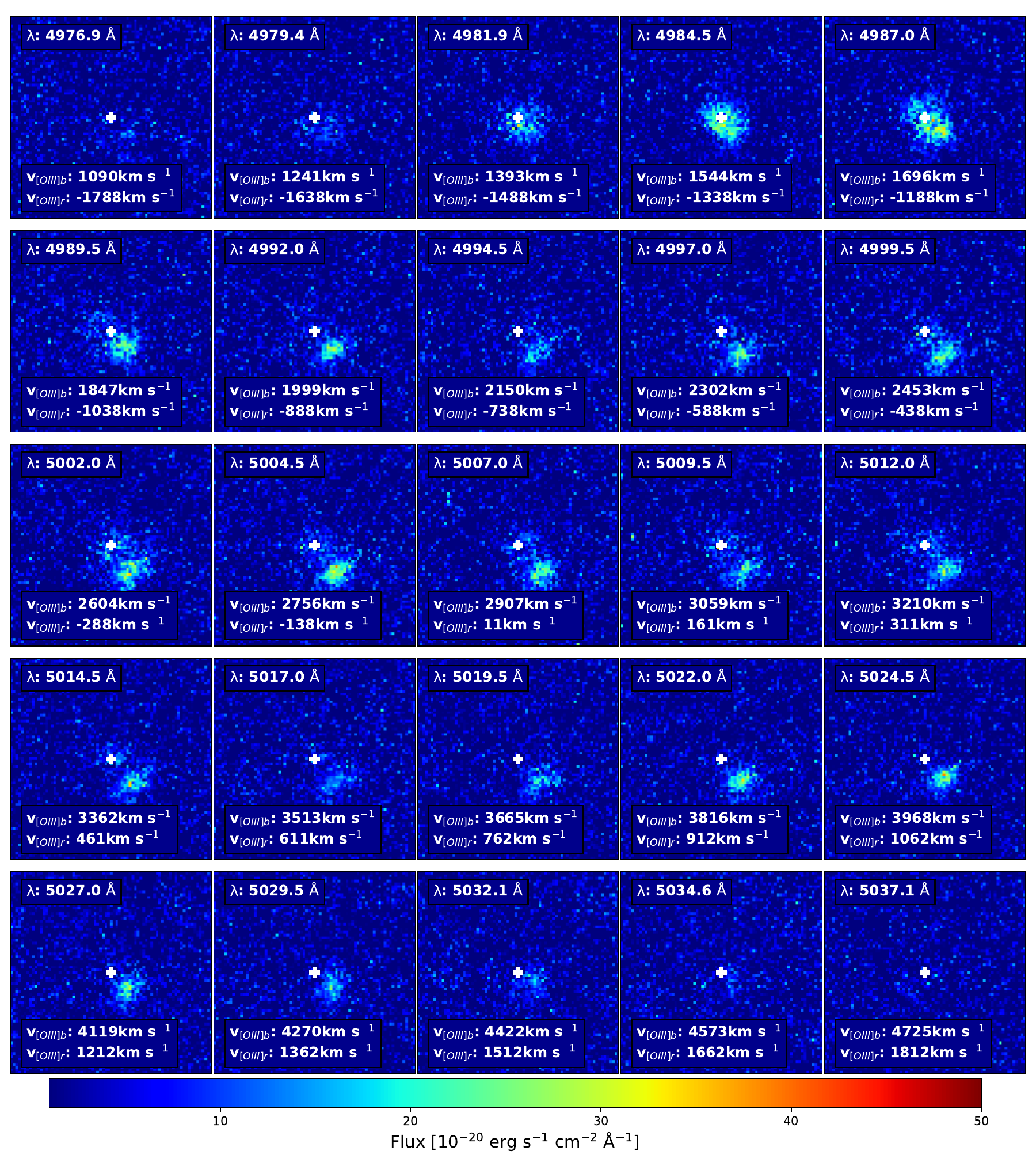}
\caption{Channel maps for [O{\sc iii}]\,$\rm\lambda$\,4059\,{\AA} ([O{\sc iii}]b) and $\rm\lambda$\,5007\,{\AA} ([O{\sc iii}]r) emission lines.}
\label{fig:appendix_2}
\end{figure}

\end{appendix}
\end{document}